\documentclass[prd,aps,floats,twocolumn,floatfix,showpacs]{revtex4}
\usepackage{graphicx,multirow,tabularx}
\usepackage{amsmath,amsthm,amssymb,mathtools,tensor,physics,slashed}
\usepackage{davitex} 

\newcommand{\dual}[1]{\prescript{*\!\!}{}{#1}}

\usepackage{hyperref}
\hypersetup{
  colorlinks=true,
  linkcolor=black,
  citecolor=black,
  urlcolor=black,
}

\usepackage{hyperref}
\hypersetup{
  colorlinks=true,
  linkcolor=black,
  citecolor=black,
  urlcolor=black,
}

\begin{document}

\title{Topological charge and cooling scales in pure SU(2)
       lattice gauge theory}
\author{Bernd A.\ Berg and David A.\ Clarke}
\affiliation{Department of Physics, Florida State University, 
             Tallahassee, FL 32306-4350, USA} 
\date{\today}

\begin{abstract}
Using Monte Carlo simulations with overrelaxation, we have equilibrated
lattices up to $\beta=2.928$, size $60^4$, for pure SU(2) lattice gauge 
theory with the Wilson action. We calculate topological charges with the 
standard cooling method and find that they become more reliable with 
increasing $\beta$ values and lattice sizes. Continuum limit estimates 
of the topological susceptibility $\chi$ are obtained of which we favor 
$\chi^{1/4}/T_c=0.643\,(12)$, where $T_c$ is the SU(2) deconfinement
temperature. Differences between cooling length scales in different 
topological sectors turn out to be too small to be detectable within 
our statistical errors.
\end{abstract} 

\maketitle

\section{Introduction} \label{sec:intro}

Since L\"uscher proposed the gradient flow method \cite{L10}, the topic 
of scale setting has received increased attention. See, for instance,
the review~\cite{So13}. In~\cite{Bo14} Bonati and D'Elia suggested 
replacing the gradient flow by the computationally more efficient 
standard cooling flow \cite{BB81} and supported this idea with 
numerical evidence for pure SU(3) lattice gauge theory (LGT). In 
a recent large statistics study of pure SU(2) LGT \cite{BC17}, we 
investigated the approach to the continuum limit for six gradient 
and six cooling scales. They are distinguished by the use of three 
different energy operators and two different ways of setting the initial 
scaling to agree with the deconfining scale on small lattices. We
studied systematic errors of scale setting which, although they are only 
about 1\% for our largest lattices at $\beta\approx 2.9$ (2\% at $\beta
\approx 2.6$), dominate the statistical errors. Quantitatively gradient 
and cooling scales worked equally well, with differences between the 
six scales within the cooling and within the gradient group larger than 
differences between corresponding scales of the two groups. See 
\cite{BC17a} for a summary. 

Using cooling we also calculated the topological charge $Q$ on each 
of our configurations and showed that our charges of subsequent
configurations are statistically independent. This was only done
for the cooling flow as it takes less than 1/34 of the CPU time needed
for the corresponding gradient flow, while the equivalence of these
scales was already demonstrated in \cite{BC17}. Here we supplement 
our previous publication by presenting details of our calculations 
of $Q$, and adding a considerable number of additional lattices at 
large $\beta$ so that we can estimate finite size corrections of 
the topological susceptibility, and come up with a continuum limit 
extrapolation.

We investigate whether there are noticeable differences in cooling 
scales when we restrict them to fixed topological sectors. Although 
fixed topological sectors imply for local operators only
a bias of order $1/V$ \cite{Br03,Ao07}, getting trapped in a
topological charge sector has often been a reason of concern. For 
instance, L\"uscher and Schaefer \cite{LS11} proposed to bypass the 
problem by imposing open boundary conditions in one of the lattice 
directions.  Recently L\"uscher \cite{L17} emphasized that master-field 
configurations on very large lattices would alleviate topological 
freezing. We find that the lattices used in our SU(2) investigation 
are so large that the $1/V$ effects due to topological freezing are 
swallowed by statistical errors.

In the next section we discuss our data for the topological charge. 
In section~\ref{sec:topsus} we present cooling scales and our 
continuum extrapolation of the topological susceptibility. In 
section~\ref{sec:scales} we search for correlations of topological 
charge sectors with differences in the considered cooling scales. 
Summary and conclusions are given in the final section~\ref{sec:sum}.

\section{Topological charge} \label{sec:topchare}

The continuum equation of the topological charge,
\begin{equation}\label{eq:Q}
  Q=\frac{g^2}{16\pi^2}\int d^4x\Tr\dual{F_{\mu\nu}}F^{\mu\nu}\,,
\end{equation}
where $\dual{F}$ is the dual field strength tensor, translates on
the lattice to the discretization
\begin{equation} \label{QL}
  Q_L=\sum_n q_L(n)\,,
\end{equation}
where the sum is over all lattice sites and
\begin{equation}
  q_L(n) = -\frac{1}{2^9\pi^2}\sum\limits_{\mu\nu\rho\sigma=\pm 1}^{\pm 4}
         \tilde{\epsilon}_{\mu\nu\rho\sigma}
         \Tr U^\Box_{\mu\nu}(n)U^\Box_{\rho\sigma}(n)\,.
\end{equation}
Here $\tilde{\epsilon}=\epsilon$ for positive indices while 
$\tilde{\epsilon}_{\mu\nu\rho\sigma}=
  -\tilde{\epsilon}_{(-\mu)\nu\rho\sigma}$ for negative indices.

\begin{table*}[th] \centering \caption{Overview of our 
largest $N^4$ lattices at fixed $\beta$ values.}
\begin{tabularx}{\linewidth}{L|C|C|C|C|C|C|R} \hline\hline
$N$ & $\beta$& $N_L^{100}$& $N_L^{1000}$& $N_L^{2048}$&
$|Q^{2048}_{\max}(1000)|$& \% stable& $|Q^{2048}_{\max}(2048)|$\\ 
\hline
16& 2.300 & 1.202 & 1.178 & 1.155 &  6& 61.7 &  3 \\
28& 2.430 & 1.258 & 1.128 & 1.129 & 15& 60.9 & 13 \\
28& 2.510 & 1.148 & 1.127 & 1.124 & 14& 66.4 & 10 \\
40& 2.574 & 1.159 & 1.117 & 1.113 & 17& 58.6 & 16 \\
40& 2.620 & 1.135 & 1.111 & 1.110 & 13& 78.1 & 12 \\
40& 2.670 & 1.131 & 1.110 & 1.108 & 10& 83.6 & 10 \\
40& 2.710 & 1.131 & 1.107 & 1.105 &  7& 87.5 &  7 \\
40& 2.751 & 1.113 & 1.108 & 1.108 &  8& 94.5 &  8 \\
44& 2.816 & 1.111 & 1.105 & 1.101 &  7& 89.1 &  7 \\
52& 2.875 & 1.112 & 1.100 & 1.098 &  7& 96.9 &  6 \\
60& 2.928 & 1.106 & 1.107 & 1.097 &  5& 96.1 &  5 \\
\hline\hline
\end{tabularx} \label{tab:sizes} \end{table*}

Measurements of this quantity on MC-generated lattice configurations 
suffer from lattice artifacts, which we suppressed by cooling. A SU(2) 
cooling step minimizes the action locally by replacing a link variable 
$U_\mu(x)$ by a function of the staple matrix $U_\mu^\sqcup(x)$:
\begin{equation}\label{eq:cool}
  U_\mu(x)\to U'_\mu(x)\equiv \frac{U^\sqcup_\mu(x)}
            {\sqrt{\det U_\mu^\sqcup(x)}}\,.
\end{equation}
After sufficiently many cooling sweeps one may reach (and does on large
enough lattices)  metastable configurations to which a topological 
charge can be assigned. Picking a suitable number $m_c$ of cooling
sweeps, the obtained charge values still suffer from discretization 
errors, which can be absorbed by multiplicative normalization constants 
$N_L$, replacing $Q_L^{m_c}$ by
\begin{equation} \label{Q}
  Q^{m_c}=N_L^{m_c}\,Q_L^{m_c}\,,~~~Q_0^{m_c}={\rm nint}(Q^{m_c})\,,
\end{equation}
where nint stands for nearest integer and we calculate the constants
$N_L^{m_c}$ following the procedure most clearly explained in 
Ref.~\cite{Bo14} and there attributed to \cite{DD02}. We minimize 
the equation
\begin{equation} \label{mini}
  \sum_{\rm conf} \{ N_L^{m_c}\,Q_L^{m_c}(\text{conf}) -
  \text{nint}[N_L^{m_c}\,Q_L^{m_c}(\rm conf)] \}^2\,,
\end{equation}
where the sum is over all configurations for a fixed lattice size 
and $\beta$ value. The integer values $Q_0^{m_c}$ protect the thus 
defined topological charge against renormalization.

All our lattices are of size $N^4$. Table~\ref{tab:sizes} gives an 
overview of our largest lattices at the $\beta$ values for which we 
calculated the topological charge distribution. For each parameter 
value we generated 128 configurations separated by a sufficiently 
large number of MCOR sweeps so that they are effectively statistically 
independent. Each MCOR update consists of one heatbath followed by two 
overrelaxation updates. For lattice sizes up to $52^4$ the statistics 
is the one of Ref.~\cite{BC17}. For our new, largest lattice, $60^4$ 
at $\beta=2.928$, lattice configurations are separated by $3\times 
2^{12}$ MCOR sweeps after $2^{15}$ sweeps for equilibration. 

\begin{figure*}[ht]
\begin{minipage}{\columnwidth}
\caption{$16^4$, $\beta=2.3$: Cooling time series $Q^{2048}(i_c)$.}
\label{fig:c16b2p3}
\includegraphics[width=0.9\columnwidth]{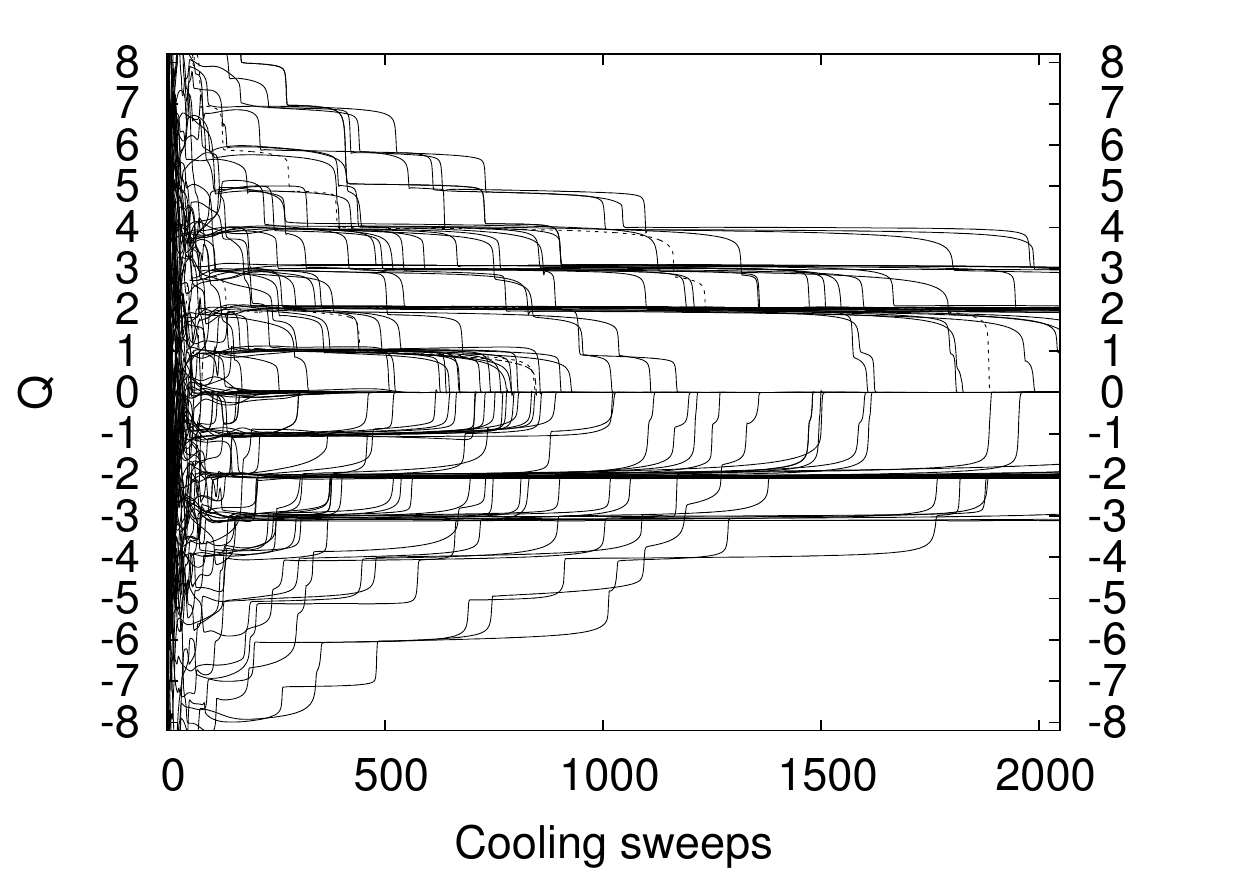}
\caption{$28^4$, $\beta=2.51$: Cooling time series $Q^{2048}(i_c)$.}
\label{fig:c28b2p51}
\includegraphics[width=0.9\columnwidth]{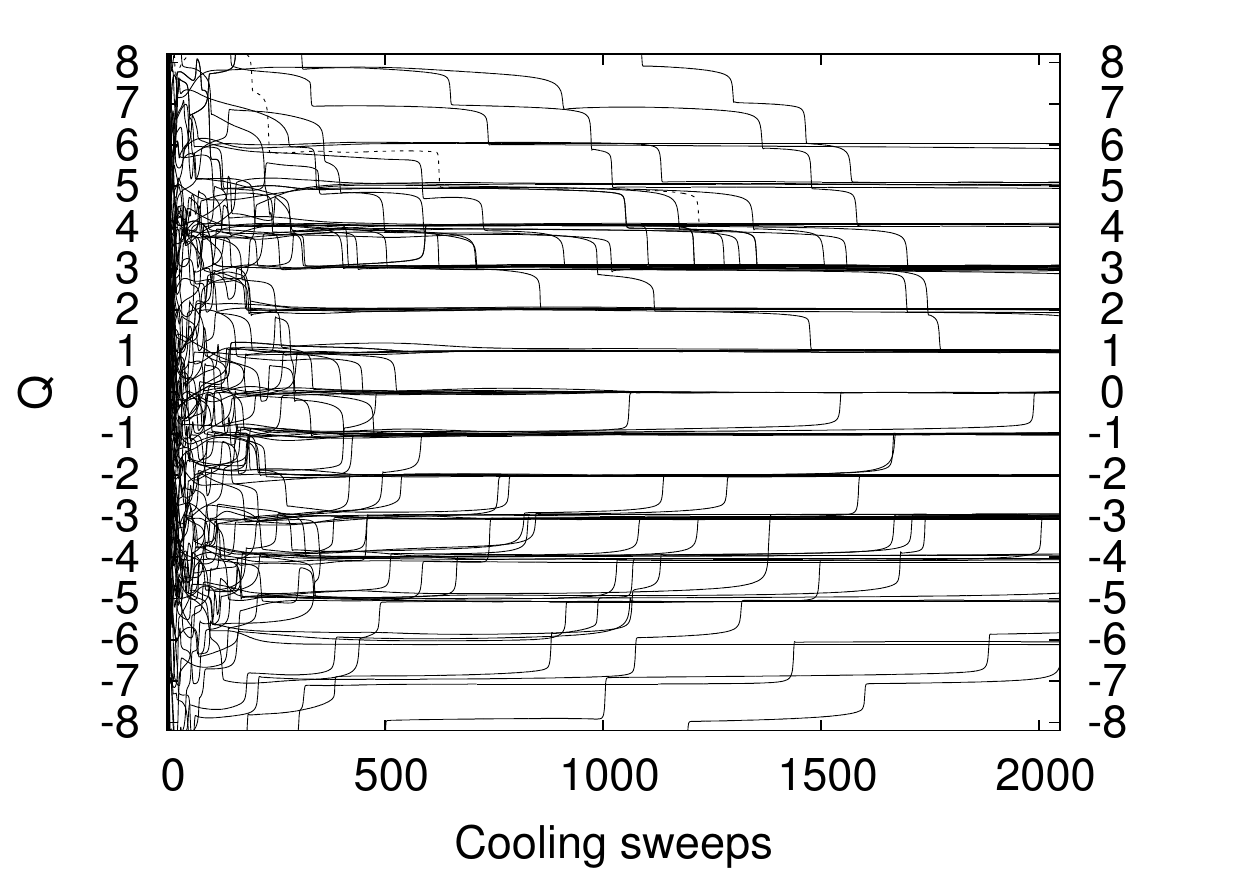}
\end{minipage}
\begin{minipage}{\columnwidth}
\caption{$40^4$, $\beta=2.751$: Cooling time series $Q^{2048}(i_c)$.}
\label{fig:c40b2p751}
\includegraphics[width=0.9\columnwidth]{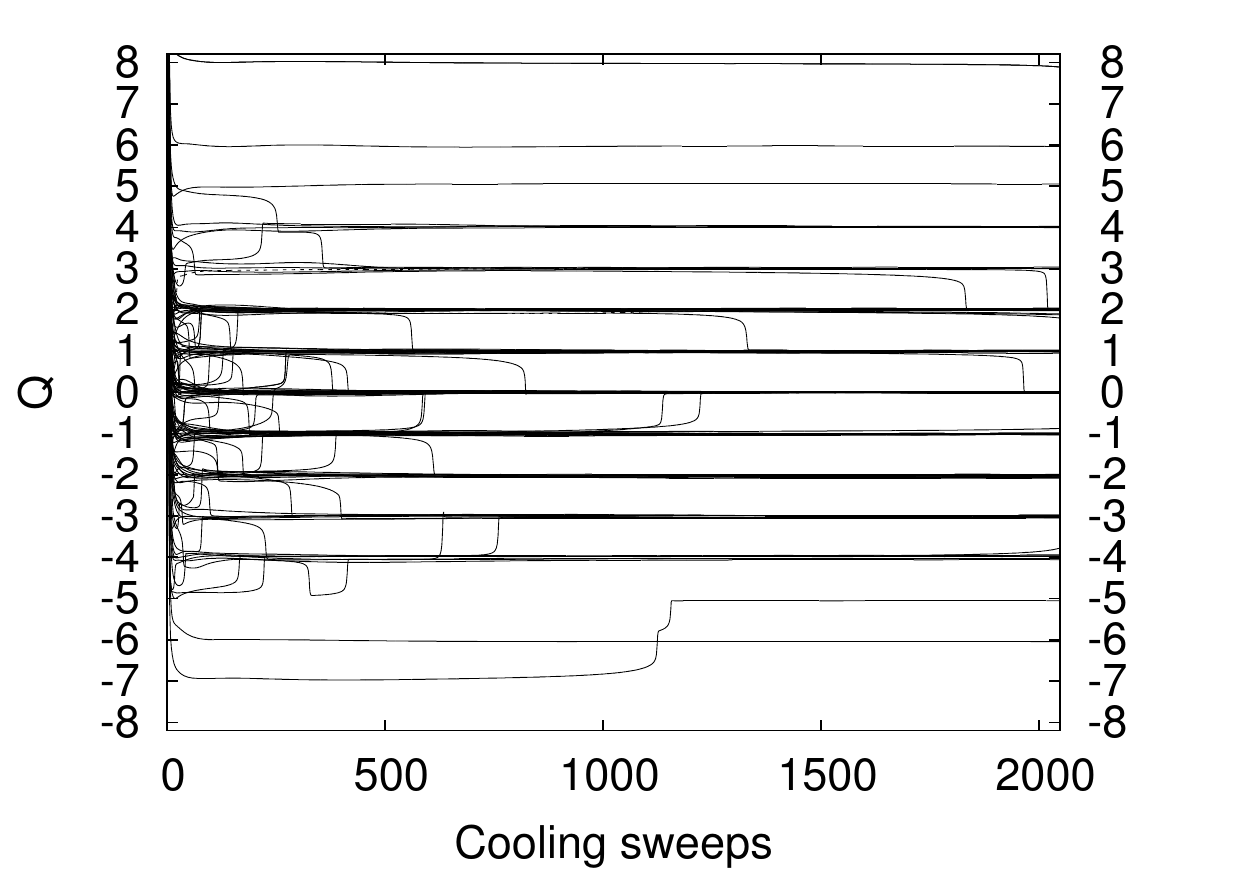}
\caption{$60^4$, $\beta=2.928$: Cooling time series $Q^{2048}(i_c)$.}
\label{fig:c60b2p928}
\includegraphics[width=0.9\columnwidth]{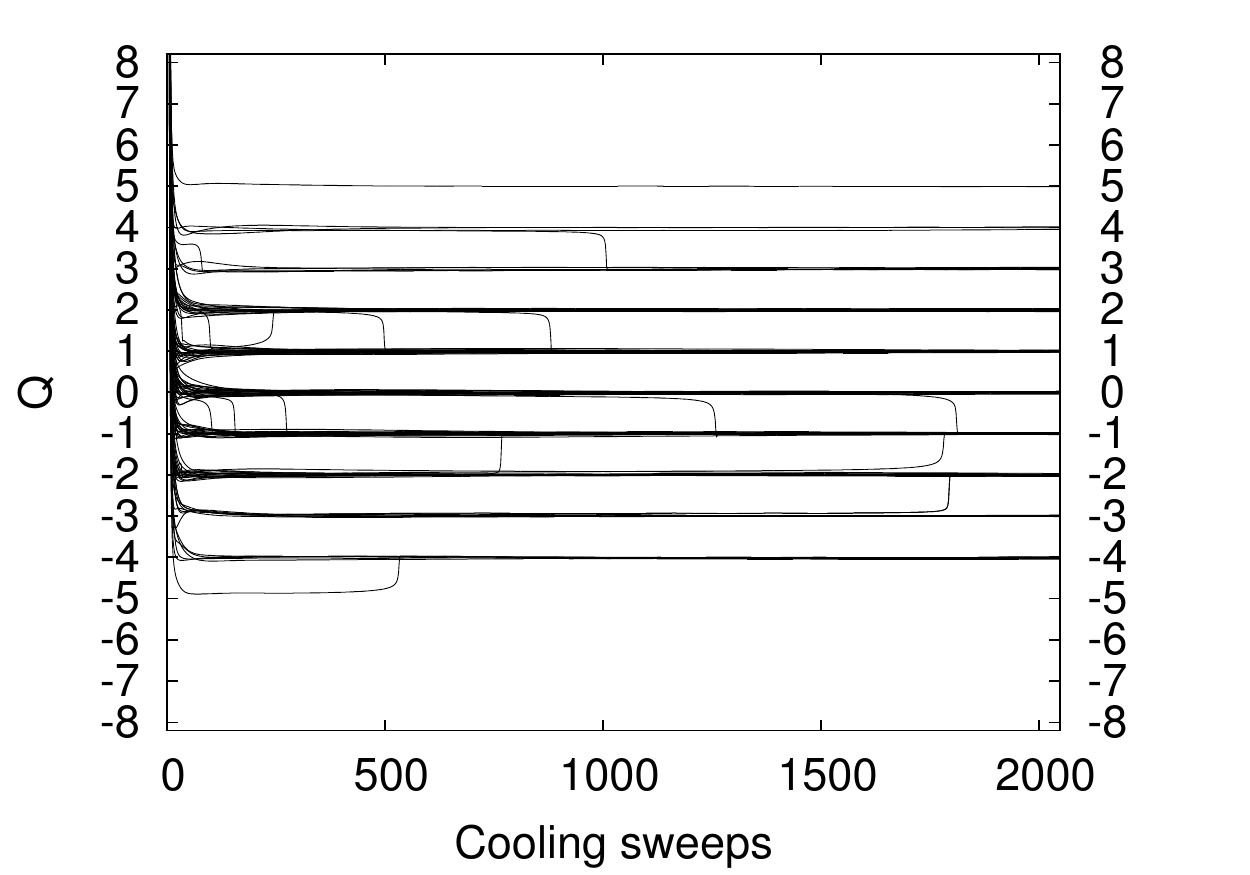}
\end{minipage}
\end{figure*}

On each lattice configuration we performed 2048 cooling sweeps and 
applied the minimization \eqref{mini} with the charges defined at 
$m_c=100$, $m_c=1000$, and $m_c=2048$. The corresponding multiplicative 
constants $N_L^{m_c}$ amount to corrections in the range from up to 26\% 
down to about 10\% for our largest lattices and $\beta$ values, where 
there is also little $m_c$ dependence of $N_L^{m_c}$. Subsequently, we 
considered plots of the $3\times 128$ time series for the topological 
charge that we created for the different $m_c$ values. For $m_c=2048$ 
examples of these plots for increasing $\beta$ values and lattice sizes 
are shown in Figs.~\ref{fig:c16b2p3} to~\ref{fig:c60b2p928}. We plot 
$Q^{2048}(i_c)$, $i_c$ number of cooling sweeps, instead of the integer 
valued charges $Q_0^{2048}(i_c)$, because the latter would obscure how 
good the mapping on integer values really is. Apart from that, using 
the integer values $Q_0^{m_c}(i_c)$ in our subsequent analysis would 
lead to the same conclusions.

As discussed in \cite{Vi09}, when approaching the continuum limit the 
topological charge has to be defined at a {\it fixed}, large enough
number $n_c$ of cooling sweeps. This number can agree with the number 
$m_c$ used for the minimization \eqref{mini}, but needs not necessarily
be identical. So our charges have two labels: 
\begin{equation}
  Q = Q^{m_c}(n_c)\,.
\end{equation}
For our first two figures a good choice of $n_c$ appears not to 
exist, because there are a considerable number of transitions between 
topological sectors over the entire times series range considered, 
and for the $16^4$ lattice the ultimate topological sector $Q=0$ 
is approached all the way. In contrast to that we find for 
Figs.~\ref{fig:c40b2p751} and~\ref{fig:c60b2p928} over a large range 
of $n_c$ values, certainly including $n_c=1000$, only few transitions.
Also the removal of dislocations by an initial number of cooling of 
sweeps becomes easier for increasing $\beta$. Metastable configurations 
are not only more stable than at lower $\beta$ values, but are also 
reached earlier. In the \% stable column of Table~\ref{tab:sizes} we 
report the stability of charge sectors under the next 1048 cooling 
sweeps after $n_c=1000$. Starting from about $\beta=2.574$ we see, up 
to statistical fluctuations, a gradually improving trend with increasing
$\beta$. If one desires that roughly 90\% of configurations are 
metastable, we must require $\beta\gtrsim2.75$ and lattices large enough 
to accommodate physical instantons (their size increases proportionally
to our length scales of \cite{BC17}, to which the largest lattice sizes 
are already adjusted).

The number $n_c=1000$ is considerably larger than what we would 
have expected from previous literature. For instance, in Fig.~3 
of \cite{Bo14} the topological charge on a $20^4$ SU(3) lattice 
at $\beta=6.2$ is defined after 21 cooling sweeps. This led us 
in \cite{BC17} to work with $n_c=100$ to define $Q^{n_c}$. For 
the purpose of checking the statistical independence of our 
configurations this is still sufficient, because including some 
dislocations adds only some statistical noise to the charge 
correlations. Early SU(2) investigations were performed for 
such small $\beta$ values and lattice sizes \cite{Te86,Il86} 
that only qualitative insights could be obtained, as already 
noted in the paper by Teper~\cite{Te86}. 

We checked that our $Q^{2048}(1000)$ charges are statistically 
independent, and that their charge distribution is symmetric under 
$Q^{2048}(1000)\to -Q^{2048}(1000)$ within statistical errors. For 
the lattices of Table~\ref{tab:sizes} histograms of $|Q^{2048}(1000)|$ 
are compiled in Table~\ref{tab:hist}. Table~\ref{tab:sizes} also 
compares the largest value of $|Q^{2048}(1000)|$ with the largest value 
of $|Q^{2048}(2048)|$, and the $|Q^{2048}_{\max}|$ values become quite 
stable for $\beta\ge2.574$.

\begin{table}[th] \centering
\caption{Histograms of $|Q^{2048}(1000)|$ for the $\beta$ values 
and lattices of Table~\ref{tab:sizes}.}
\begin{tabularx}{\columnwidth}{l|C|C|C|C|C|C|C|C|C|C|C|C|C|C|C|C|R} 
\hline\hline
$\beta$& 0& 1& 2& 3& 4& 5& 6& 7& 8& 9&10&11&12&13&14&15&17\\ \hline
 2.300 &57& 4&36&20& 6& 4& 1& 0& 0& 0& 0& 0& 0& 0& 0& 0& 0\\
 2.430 & 6&22&15&15&22&10&10& 5& 7& 7& 2& 2& 1& 2& 0& 2& 0\\
 2.510 &11&21&17&23&19&17& 7& 3& 1& 5& 2& 1& 0& 0& 1& 0& 0\\
 2.574 &11&12&19&14&14&12&10&12& 5& 2& 5& 6& 3& 0& 2& 0& 1\\
 2.620 &13&18&23&19&13&13& 7& 5& 2& 3& 5& 3& 3& 1& 0& 0& 0\\
 2.670 &12&28&31&11&15&12& 9& 3& 3& 3& 1& 0& 0& 0& 0& 0& 0\\
 2.710 &20&30&33&23&11& 7& 3& 1& 0& 0& 0& 0& 0& 0& 0& 0& 0\\
 2.751 &28&37&31&16&11& 1& 2& 1& 1& 0& 0& 0& 0& 0& 0& 0& 0\\
 2.816 &24&42&32&18& 9& 1& 1& 1& 0& 0& 0& 0& 0& 0& 0& 0& 0\\
 2.875 &29&40&27&24& 5& 2& 0& 1& 0& 0& 0& 0& 0& 0& 0& 0& 0\\
 2.928 &26&49&30&12&10& 1& 0& 0& 0& 0& 0& 0& 0& 0& 0& 0& 0\\
\hline\hline
\end{tabularx} \label{tab:hist} \end{table}

\section{Cooling scales and topological susceptibility 
\label{sec:topsus}}

\begin{table}[th] 
\centering 
\caption{\label{tab_ct01}{Cooling length scales for the $y^{01}_i$ 
set. The * denotes lattices that are too small to be used for finite 
size fits. TVNR stands for ``target value not reached''.}} \smallskip
\begin{tabularx}{\columnwidth}{L|C|C|C|R} \hline\hline
$\beta$& $N$&$L_7=s_0^{01}$&$L_8=s_1^{01}$&$L_9=s_4^{01}$\\ \hline
2.300& $16$ & 1.3433(24)& 1.3385(23)& 1.2575(74) \\ 
2.430& $28$ & 2.0892(28)& 2.0707(28)& 1.9446(95) \\ 
2.510& $28$ & 2.7522(68)& 2.7267(66)& 2.548(15) \\ 
2.574*& $16$ & 3.512(48)& 3.478(47)& 3.309(48) \\ 
2.574& $28$ & 3.422(13)& 3.390(13)& 3.168(18) \\ 
2.574& $40$ & 3.4048(69)& 3.3730(67)& 3.137(17) \\ 
2.620*& $16$ & 4.55(14)& 4.50(26)& 4.28(12) \\ 
2.620& $28$ & 3.9752(19)& 3.915(19)& 3.690(24) \\
2.620& $40$ & 3.9509(95)& 3.913(93)& 3.645(22) \\ 
2.670*& $16$ & 6.28(36)& 6.23(36)& 5.88(38) \\
2.670& $28$ & 4.676(32)& 4.631(31)& 4.314(39) \\
2.670& $40$ & 4.618(17)& 4.574(16)& 4.298(26) \\ 
2.710*& $16$ & 8.03(76)& 7.96(80)& 7.67(1.3)\\ 
2.710& $28$ & 5.232(41)& 5.184(40)& 4.829(47) \\ 
2.710& $40$ & 5.203(21)& 5.154(21)& 4.794(28) \\ 
2.751*& $16$& TVNR     & TVNR     & TVNR      \\ 
2.751& $28$ & 5.880(82)& 5.824(78)& 5.487(74) \\ 
2.751& $40$ & 5.913(32)& 5.857(32)& 5.434(40) \\ 
2.816& $28$ & 8.247(27)& 8.167(26)& 7.561(25) \\
2.816& $40$ & 7.089(58)& 7.021(58)& 6.517(68) \\
2.816& $44$ & 7.105(45)& 7.039(45)& 6.511(55) \\ 
2.875*& $28$ & 12.84(84)& 12.06(83)&11.70(84) \\ 
2.875& $40$ & 8.55(11)& 8.464(10)& 7.885(97) \\
2.875& $44$ & 8.637(93)& 8.554(92)& 7.912(89) \\
2.875& $52$ & 8.514(60)& 8.433(59)& 7.825(68) \\ 
2.928*& $28$ &16.3(1.8)&16.2(1.8)&14.8(1.7)\\
2.928& $40$ & 10.90(30)& 10.79(29)& 9.89(27)\\
2.928& $44$ & 10.01(16)& 9.92(16)& 9.18(14)\\
2.928& $52$ & 9.940(88)& 9.846(87)& 9.112(93)\\
2.928& $60$ & 9.835(67)& 9.742(66)& 9.053(70)\\ 
\hline\hline
\end{tabularx} 
\end{table} 

\begin{table}[th] 
\centering 
\caption{\label{tab_ct02}{Cooling length scales for the $y^{02}_i$ 
set. The * denotes lattices that are too small to be used for finite
size fits. TVNR stands for ``target value not reached''. }} \smallskip
\begin{tabularx}{\columnwidth}{L|C|C|C|R} \hline\hline
$\beta$& $N$&$L_{10}=s_0^{02}$&$L_{11}=s_1^{02}$
            &$L_{12}=s_4^{02}$\\ \hline
2.300  & $16$ & 1.8307(39)& 1.8282(39)& 1.728(10) \\ 
2.430 & $28$ & 2.7317(43)& 2.7212(42)& 2.565(12) \\ 
2.510 & $28$ & 3.552(10)& 3.5371(99)& 3.315(18) \\ 
2.574& $16$ & 4.550(69)& 4.529(65)& 4.323(71) \\ 
2.574& $28$ & 4.405(20)& 4.386(29)& 4.123(25) \\ 
2.574& $40$ & 4.377(11)& 4.358(10)& 4.074(20) \\ 
2.620* & $16$ & 5.82(17)& 5.80(17)& 5.58(17) \\ 
2.620 & $28$ & 5.104(31)& 5.082(31)& 4.787(35)\\
2.620 & $40$ & 5.068(15)& 5.045(15)& 4.725(26) \\ 
2.670 & $16$ & 7.92(54)& 7.89(54)& 7.57(53) \\ 
2.670 & $28$ & 6.021(46)& 5.993(46)& 5.603(58)\\
2.670 & $40$ & 5.910(25)& 5.884(25)& 5.536(33) \\ 
2.710* & $16$ &9.88(2.3)&9.86(2.3)&9.72(2.1) \\ 
2.710 & $28$ & 6.675(58)& 6.645(57)& 6.228(67) \\ 
2.710 & $40$ & 6.656(31)& 6.626(30)& 6.188(38) \\ 
2.751*& $16$ & TVNR     & TVNR     & TVNR      \\ 
2.751& $28$ & 7.55(13)& 7.52(13)& 7.07(11) \\ 
2.751& $40$ & 7.576(46)& 7.541(46)& 7.038(54) \\
2.816& $28$ & 10.48(35)& 10.44(35)& 9.72(34) \\
2.816& $40$ & 9.076(84)& 9.034(84)& 8.426(92) \\
2.816& $44$ & 9.056(65)& 9.015(64)& 8.349(73) \\ 
2.875*& $28$ & 14.66(92)& 14.62(92)&14.26(96) \\ 
2.875& $40$ & 10.98(16)& 10.93(16)&10.21(16) \\ 
2.875& $44$ & 11.11(15)& 11.06(15)&10.29(15) \\
2.875& $52$ &10.879(87)&10.830(86)&10.122(92) \\ 
2.928*& $28$ &20.3(2.3)&20.0(2.3)&17.4(2.0)\\
2.928& $40$ &13.99(42)&13.92(42)&12.87(40) \\
2.928& $44$ &12.78(23)&12.72(23)&11.82(21) \\
2.928& $52$ &12.72(13)&12.67(13)&11.76(13) \\
2.928& $60$ &12.561(97)&12.503(96)&11.653(95) \\ 
\hline\hline
\end{tabularx} 
\end{table} 

\begin{table}[th] 
\centering        
\caption{\label{tab_sus}{Topological susceptibility
defined after 1000 and 100 cooling sweeps respectively. The * 
denotes lattices that are too small to be used. }} \smallskip
\begin{tabularx}{\columnwidth}{l|c|C|C|C|R} \hline\hline
        & & 1000         &                     & 100 &        \\ \hline
$\beta$ &$N$& $\chi^{1/4}$ & $L_{10}\,\chi^{1/4}$ 
          & $\chi^{1/4}$ & $L_{10}\,\chi^{1/4}$ \\ \hline
2.300& $16$ &0.0903(28)&0.1654(52)&0.1231(35)&0.2253(64)\\ 
2.430& $28$ &0.0834(27)&0.2276(72)&0.1023(33)&0.2790(89)\\ 
2.510& $28$ &0.0744(25)&0.2642(86)&0.0821(26)&0.2917(90)\\ 
2.574*&$16$ &0.0510(37)&0.232(16)&0.0667(21)&0.3033(82)\\ 
2.574& $28$ &0.0601(18)&0.2647(77)&0.0653(26)&0.288(11)\\ 
2.574& $40$ &0.0609(19)&0.2666(80)&0.0677(21)&0.2963(92)\\ 
2.620*&$16$ &0.0291(32)&0.169(17)&0.0562(20)&0.3272(63)\\ 
2.620& $28$ &0.0537(16)&0.2740(76)&0.0570(16)&0.2912(79)\\ 
2.620& $40$ &0.0557(19)&0.2821(93)&0.0582(19)&0.2950(94)\\ 
2.670*&$16$ &0.026(26)&0.21(21)&0.0419(25)&0.332(13)\\ 
2.670& $28$ &0.0467(15)&0.2811(81)&0.0477(15)&0.2873(83)\\ 
2.670& $40$ &0.0484(16)&0.2860(90)&0.0511(17)&0.3020(96)\\ 
2.710*& $16$&  0         & 0         &0.0345(25)&0.341(75)\\ 
2.710& $28$ &0.0444(16)&0.2966(97)&0.0460(17)&0.307(11)\\ 
2.710& $40$ &0.0404(12)&0.2692(77)&0.0416(13)&0.2772(82)\\ 
2.751& $28$ &0.0387(15)&0.2925(96)&0.0399(16)&0.3010(98)\\ 
2.751& $40$ &0.0381(15)&0.286(11)&0.0385(15)&0.290(11)\\ 
2.816& $28$ &0.0305(15)&0.3195(97)&0.0327(18)&0.343(14)\\ 
2.816& $40$ &0.0324(12)&0.294(10)&0.0328(12)&0.298(10)\\ 
2.816& $44$ &0.0332(12)&0.3010(96)&0.0336(12)&0.3045(96)\\ 
2.875*&$28$ &0.0227(16)&0.333(12)&0.0390(17)&0.3512(94)\\ 
2.875& $40$ &0.02748(89)&0.3017(87)&0.02800(96)&0.3074(93)\\ 
2.875& $44$ &0.02681(92)&0.2980(92)&0.0270(11)&0.300(11)\\ 
2.875& $52$ &0.02760(92)&0.3002(97)&0.02822(93)&0.3070(97)\\ 
2.928*&$28$ &0.0173(17)&0.345(12)&0.0173(17)&0.345(14)\\ 
2.928& $40$ &0.0235(11)&0.3287(97)&0.0235(11)&0.3286(98)\\ 
2.928& $44$ &0.02492(77)&0.3185(75)&0.02534(85)&0.3239(84)\\ 
2.928& $52$ &0.02359(81)&0.3002(94)&0.02360(80)&0.3003(93)\\ 
2.928& $60$ &0.02297(70)&0.2885(84)&0.02313(72)&0.2906(87)\\ 
\hline\hline
\end{tabularx} 
\end{table} 

To investigate the scaling behavior of the topological susceptibility,
we have considerably extended our previous statistics by adding smaller 
lattices at each $\beta$ value. Results for the cooling length scales 
are discussed in the next subsection followed by an analysis of the 
topological susceptibility in subsection~\ref{subsec:sus}.

\subsection{Cooling length scales} \label{subsec:L}

Data for the cooling length scales are compiled in Tables~\ref{tab_ct01} 
and~\ref{tab_ct02}. For the convenience of the reader we include for 
each $\beta$ value the largest lattice, although they can already be 
found in \cite{BC17}, with the exception of $60^4$ at $\beta=2.928$. 
The $28^4$ lattices at $\beta=2.620$ and $\beta=2.670$ are also from 
\cite{BC17}. All other lattices are from new simulations. For them we 
did not calculate the gradient length scales, because the gradient flow 
takes at least 34 times more CPU time than the cooling flow.

Following \cite{BC17} we use for the calculation of the length scales 
three definitions of the energy density: $E_0(t)$, $E_1(t)$, and $E_4(t)$. 
$E_0(t)$ is the Wilson action up to a constant factor, 
$E_1(t)$ is the sum of the squared Pauli matrices of the plaquette 
variables, and $E_4(t)$ is L\"uscher's \cite{L10} energy density which 
averages over the four plaquettes attached to each site $n$ in a fixed 
$\mu\nu$, $\mu\ne\nu$ plane.  The functions
\begin{eqnarray} \label{yi}
  y_i(t)\ =\ t^2\,E_i(t)\,,~~(i=0,1,4)
\end{eqnarray}
are used to set up three cooling scales by choosing appropriate fixed 
target values $y_i^0$ and performing cooling steps \eqref{eq:cool} 
until $y_i^0=(t_i^0)^2\,E_i(t_i^0)$ is reached. As a function of $\beta$, 
the observable
\begin{eqnarray} \label{s0i}
  s_i^0(\beta)\ =\ \sqrt{t_i^0(\beta)}
\end{eqnarray}
then scales like a length.

There is some ambiguity in the choice of target values. In \cite{BC17} 
they are chosen so that either (superscripts $01$) initial estimates of 
the scales $s_0^{01}$ and $s_1^{01}$ (they give almost identical values) 
agree with the deconfinement scaling from $\beta\approx 2.3$ on a 
$4\times 8^3$ lattice to $\beta\approx 2.44$ on a $6\times 12^3$ lattice, 
or so that (superscripts $02$) $s_4^{02}$ agrees. This leads to two 
possible values per energy observable, i.e., a total of six targets:
\begin{eqnarray} \label{cyi01}
  y^{01}_0&=&0.0440\,,~~y^{01}_1\ =\ 0.0430\,,~~y^{01}_4\ =\ 0.0350\,,
  ~~~~\\ \label{cyi02} 
  y^{02}_0&=&0.0822\,,~~y^{02}_1\ =\ 0.0812\,,~~y^{02}_4\ =\ 0.0656\,.
  ~~~~
\end{eqnarray}
The numeration of the scales as $L_7$ to $L_{10}$ follows the convention 
of \cite{BC17}, where $L_1$ to $L_6$ are the corresponding gradient
scales.

\subsection{Topological susceptibility} \label{subsec:sus}

At each $\beta$ and lattice size,
we calculated the topological susceptibility
\begin{eqnarray} \label{sus}
  \chi = \frac{1}{n}\frac{1}{N^4}\sum_{i=1}^{n}\left<Q_i^2\right>,
\end{eqnarray}
where the sum runs over our $n=128$ configurations at two fixed 
$(m_c,n_c)$ values,
\begin{eqnarray} \label{mcnc}
  n_c=m_c=100~~~{\rm and}~~n_c=m_c=1000.
\end{eqnarray}

Results for $\chi^{1/4}$ with jackknife error bars are given in 
Table~\ref{tab_sus}. For $\beta\to\infty$ the product of $\chi^{1/4}$ 
with one of our cooling length scales should approach a constant up to 
$a^2$ corrections ($a$ lattice spacing). As in Figs.~7, 8 and~10 of 
Ref.~\cite{BC17} we choose the $L_{10}$ length scale as our reference 
and report in Table~\ref{tab_sus} estimates of $L_{10}\,\chi^{1/4}$. The 
quantities cannot simply be obtained by multiplying the $L_{10}$ values 
of Table~\ref{tab_ct02} with the $\chi^{1/4}$ estimates of 
Table~\ref{tab_sus} and using error propagation, because their values 
come from the same configurations. Instead 128 jackknife bins were 
calculated for the product $L_{10}\,\chi^{1/4}$ at each fixed lattice 
size and $\beta$ value. The given error bars are from these jackknife 
bins.

\begin{figure}[t]
\caption{Cooling time series of $L_{10}\,\chi^{1/4}$ for 2048 cooling 
sweeps on the bottom abscissa (for the red curves) and 200 cooling 
sweeps on the top abscissa (for the blue curves). Lattice parameters 
are given in the text.} \label{fig:s1o4}
\includegraphics[width=\columnwidth]{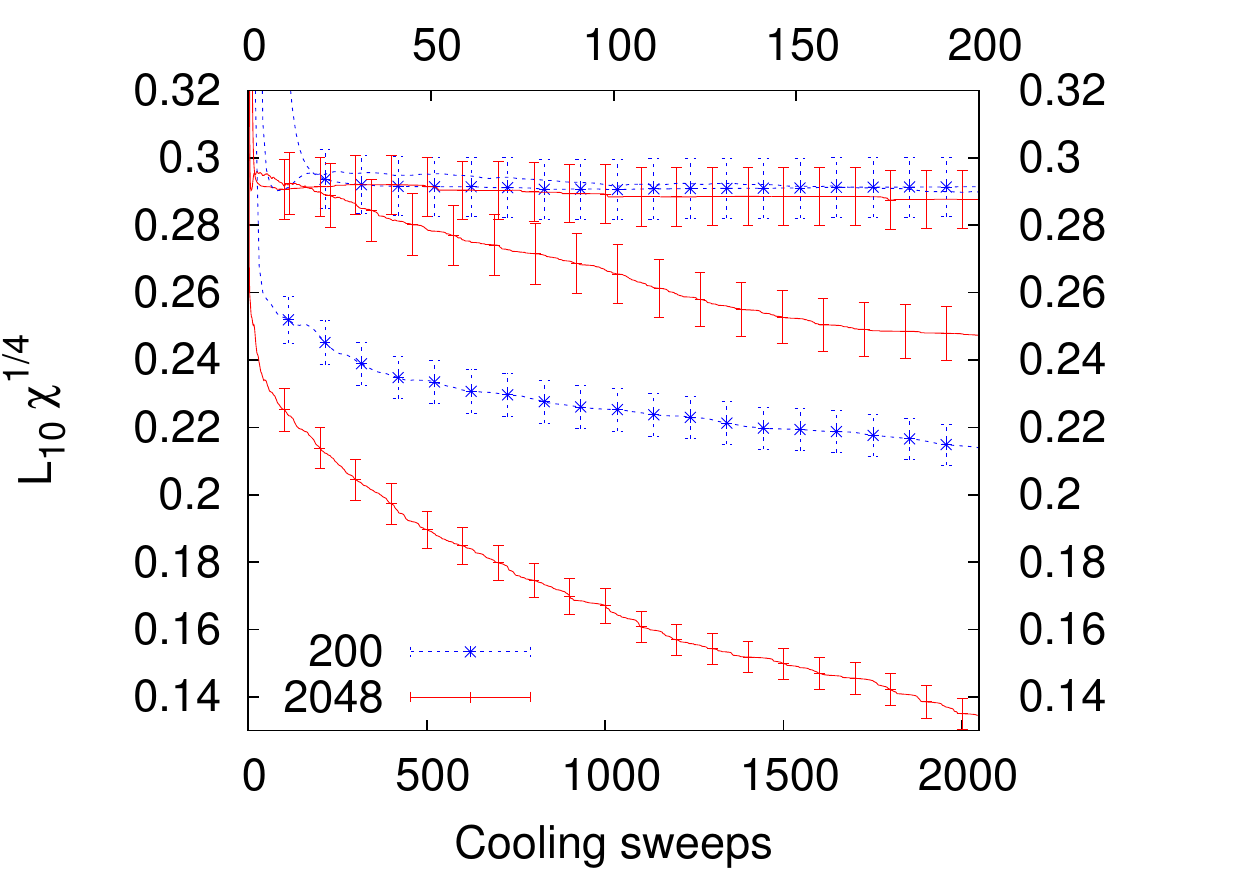}
\end{figure}

Figure~\ref{fig:s1o4} shows the time evolution of $L_{10}\,\chi^{1/4}$
for the same lattice sizes and $\beta$ values that we used to exhibit 
the time evolution of the topological charge in 
Figs.~\ref{fig:c16b2p3} to~\ref{fig:c60b2p928}. There are three 
almost constant lines near the top of Fig.~\ref{fig:s1o4}: A red 
line and two blue lines that fall practically on top of one 
another. The red line and one of the blue lines belong to to the $60^4$ 
lattice at $\beta=2.928$ used for Fig.~\ref{fig:c60b2p928}. Red curves
are to be read using the bottom abscissa with error bars plotted every
100 cooling sweeps, while blue curves are to be read using the top 
abscissa with error bars plotted every 10 sweeps. For the $60^4$ the
blue line stays constant and the red line continues this out to 2048 
cooling sweeps.

We do not include the cooling time series for the $40^4$ lattice at
$\beta=2.751$ in Fig.~\ref{fig:s1o4}, because they fall on top of the 
time series of the $60^4$ lattice at $\beta=2.928$.

Next we consider $L_{10}\,\chi^{1/4}$ from the $28^4$ lattice at $\beta
=2.51$ under cooling, given by the topmost, decreasing red curve.
As one may have expected from the time evolution of the topological 
charge in Fig.~\ref{fig:c28b2p51}, its susceptibility decreases 
monotonically. However, the behavior of the scale during the first 
200 cooling sweeps comes as a surprise. It is given by a second blue 
line that falls almost on top of the blue line for $L_{10}\,\chi^{1/4}$ 
from the $60^4$ lattice at $\beta=2.928$. In Fig.~\ref{fig:c28b2p51} 
there are many transitions between topological sectors in this range. 
So, an almost constant topological susceptibility is only possible 
when the transitions that increase the topological charge are, within 
statistical errors, matched by those that decrease it. An enhancement 
of the first 200 cooling sweeps of Fig.~\ref{fig:c28b2p51} confirms 
this scenario. Notably, even in the range of less than 200 cooling 
sweeps $L_{10}\chi^{1/4}$ scales then already very well as a constant 
all the way from $\beta=2.51$ to $\beta=2.928$.

The lowest blue and red curves correspond to the $16^4$ lattice at 
$\beta=2.3$ that was used for Fig.~\ref{fig:c16b2p3}. Both curves 
are now monotonically decreasing and demonstrate that $\beta=2.3$ 
is too small to provide a reliable estimate of the topological 
susceptibility.

\begin{figure}[t]
\caption{Cooling trajectories (with error bars) at $\beta=2.751$ for 
different lattice sizes. The dashed line indicates the $L_{10}$
target value $y_0^{02}$ (\ref{cyi02}).}
\includegraphics[width=\columnwidth]{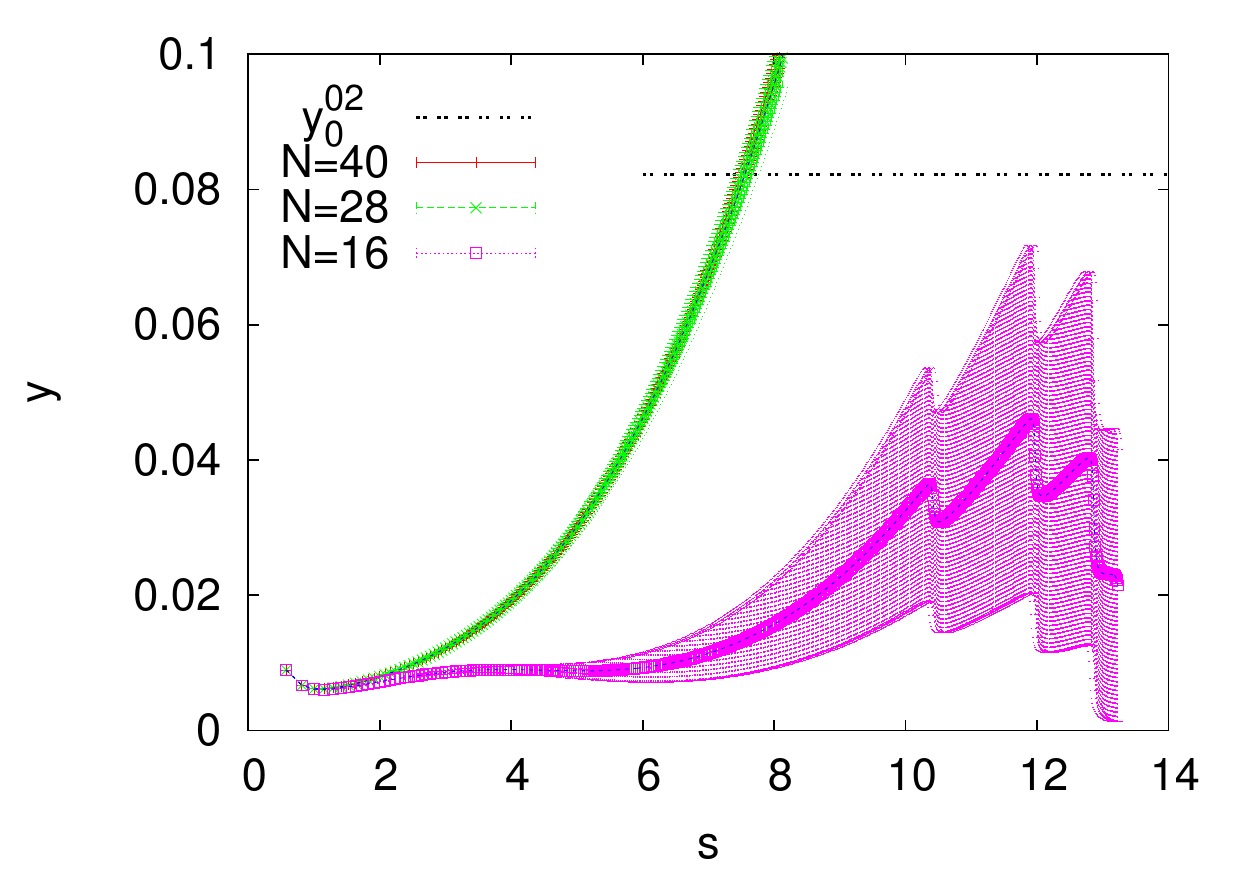}
\label{fig:cool}
\end{figure}

Let us now discuss scaling and continuum limit extrapolation of 
$\chi^{1/4}$. For this purpose we combine the results 
for $L_{10}\,\chi^{1/4}$ at fixed $\beta$ but different $N$ using
two-parameter fits
\begin{equation}\label{eq:fss}
L_{10}\,\chi^{1/4}(\beta,N)=a_1+\frac{a_2}{N^4},
\end{equation}
where $a_1$ serves as an estimator for $L_{10}\,\chi^{1/4}(\beta)$.

The lattices with * in the first column of Tables~\ref{tab_ct01}, 
\ref{tab_ct02} and~\ref{tab_sus} turned out to be too small to deliver 
reliable data and are therefore not included in these fits. 
\begin{figure}[t]
\caption{Scaling of $L_{10}\,\chi^{1/4}$. The top (bottom) set of fits 
uses the top (bottom) abscissa and right (left) ordinate. }
\includegraphics[width=\columnwidth]{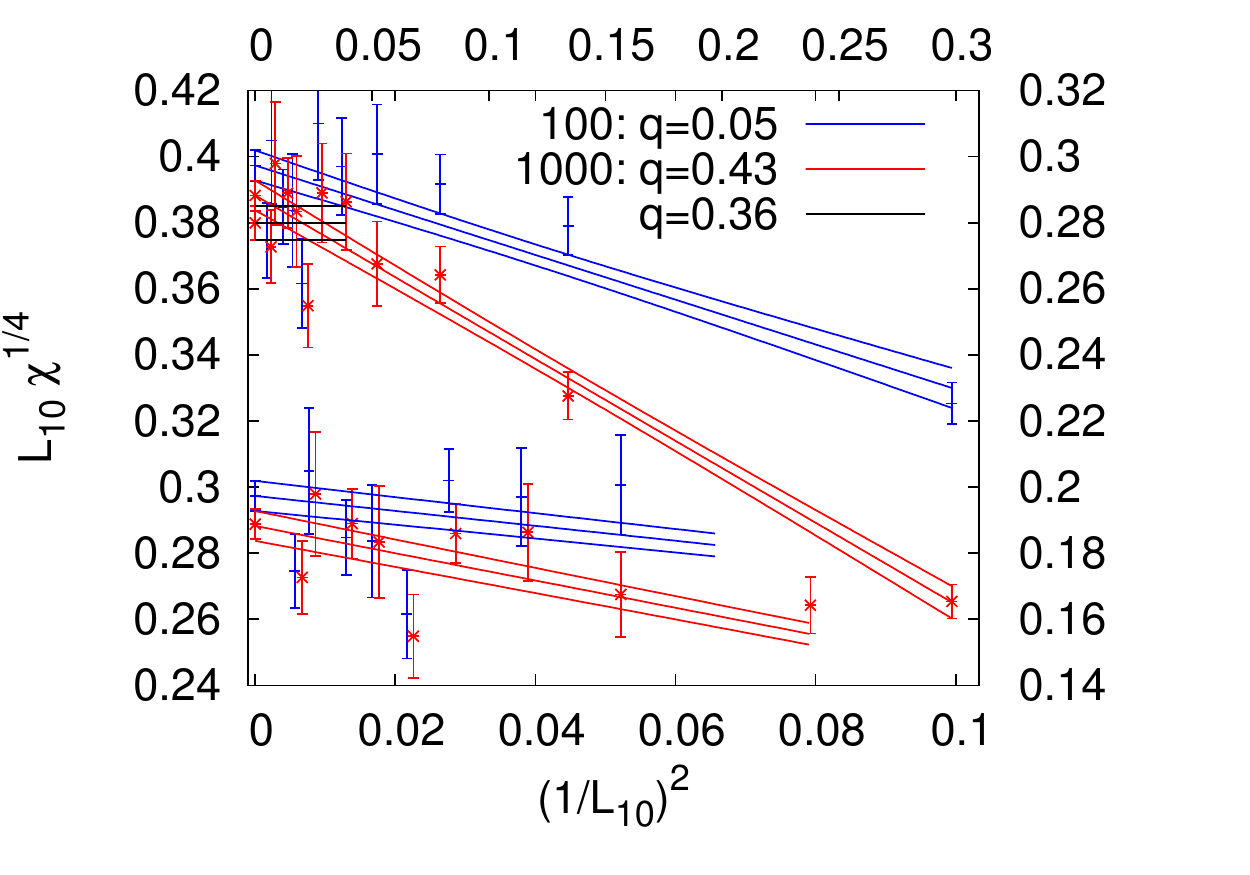}
\label{fig:BAV}
\end{figure}
For instance, as reported in Table~\ref{tab_sus}, at $\beta=2.71$ the 
topological susceptibility is zero at $n_c=1000$ for the $16^4$ lattice, 
implying that the topological charge is zero on each of our 128 
configurations. Also the cooling scale breaks down at high $\beta$ 
values when the lattice sizes are too small. For the $16^4$ lattice this 
happens for $\beta\ge 2.751$, and is illustrated in Fig.~\ref{fig:cool}. 
The trajectories for the $40^4$ and the $28^4$ lattice fall nicely on
top of one another, so that in the figure only the color of the second 
drawn trajectory is left over. However for the $16^4$ lattice, the 
trajectory fails to take off, so that the $y_0^{02}=0.0822$ target
value (\ref{cyi02}) for $L_{10}$ is never reached.

Carrying out the fit~\eqref{eq:fss} on the remaining lattices yields 
results consistent with the fitting form. In particular, in order of 
$\beta=2.928$, 2.875 and 2.816, the goodness of fit is $q=0.92$, 0.78 
and 0.48 for $n_c=1000$, and $q=0.72$, 0.57 and 0.40 for $n_c=100$. For 
$\beta=2.751$, 2.71, 2.67, 2.62 and 2.574 we performed two-parameter 
fits with only 2 lattices so that there are no $q$-values to report. 
For $\beta=2.51$, 2.43 and 2.3 the result from the single lattice 
listed in Table~\ref{tab_sus} is taken in each case.

In Fig.~\ref{fig:BAV} we show different fits of the thus obtained data.
Using the $L_{10}\,\chi^{1/4}$ estimates down to $\beta=2.3$, linear 
fits to  $a^2$ scaling corrections given by $1/(L_{10})^2$ are shown 
in the upper part of the figure along with their error bar ranges, while 
the lower part shows an enhancement. The continuum limit extrapolations 
are
\begin{eqnarray} \label{L10s1o4L1000}
  L_{10}\,\chi^{1/4} &=& 0.2882\ (46),~q=0.43~{\rm for}~n_c=1000,
  ~~\\ \label{L10s1o4L0100}
  L_{10}\,\chi^{1/4} &=& 0.2961\ (49),~q=0.05~{\rm for}~n_c=100.
\end{eqnarray}

Although the fits to $a^2$ scaling corrections work well, one may question
whether the $L_{10}\,\chi^{1/4}$ results at $\beta=2.3$ and 2.43 and to 
some extent also at $\beta=2.51$ and 2.574 are really reliable. In short,
one could argue in favor or against taking out all $\beta$ values for
which the susceptibility after $n_c=100$ cooling sweeps is significantly
larger than after $n_c=1000$ cooling sweeps. Taking them out and fitting 
the remaining points to $L_{10}\,\chi^{1/4}=constant$, one obtains the 
estimates
\begin{eqnarray} \label{L10s1o4C1000}
  L_{10}\,\chi^{1/4} &=& 0.2799\ (51),~q=0.36~{\rm for}~n_c=1000,
  ~~~~\\ \label{L10s1o4C0100}
  L_{10}\,\chi^{1/4} &=& 0.2844\ (54),~q=0.25~{\rm for}~n_c=100.
\end{eqnarray}
To avoid overloading Fig.~\ref{fig:BAV}, the fit to a constant is only 
indicated for $n_c=1000$ in the upper part of the figure.

\begin{table}[t] \centering 
\caption{\label{tab_s1o4}{Estimates of the topological susceptibility 
in units of the string tension $\sqrt{\sigma}$. }} 
\smallskip
\begin{tabularx}{\columnwidth}{l|C|C|R} \hline\hline
[Reference] (year)&$\chi^{1/4}/\sqrt{\sigma}$&$q_{1000}$&$q_{100}$\\ \hline
\cite{dF97} (1997)& 0.501  (45) & 0.32& 0.44  \\ 
\cite{dG97} (1997)& 0.528  (21) & 0.00& 0.01  \\ 
\cite{Al97} (1997)& 0.480  (23) & 0.32& 0.56  \\ 
\cite{Lu01} (2001)& 0.4831 (56) & 0.01& 0.09  \\ 
\cite{Lu01} (2001)& 0.4745 (63) & 0.07& 0.40  \\ 
\cite{Lu01} (2001)& 0.4742 (56) & 0.06& 0.40  \\ 
\hline\hline
\end{tabularx} \end{table} 

Averaging Eq.~(\ref{L10s1o4L1000}) with (\ref{L10s1o4C1000}), and
Eq.~(\ref{L10s1o4L0100}) with (\ref{L10s1o4C0100}), we obtain
\begin{eqnarray} \label{L10s1000}
  L_{10}\,\chi^{1/4}&=&0.2841\ (49 )~~{\rm for}~~n_c=1000\,,
  \\ \label{L10s0100}
  L_{10}\,\chi^{1/4}&=&0.2903\ (52 )~~{\rm for}~~n_c=100\,.
\end{eqnarray}

\begin{table*}[t]\centering
\caption{Cooling scales on topological sectors of our largest lattices
for $\beta\ge 2.71$.}
\begin{tabularx}{\linewidth}{L|C|C|C|C|C|C|C|R} \hline\hline
$\beta$ & $|Q^{1000}|$&$n$& $L_7$   & $L_8$     & $L_9$          
                  	& $L_{10}$  & $L_{11}$  & $L_{12}$  \\ \hline
2.928	& 0 	& 26    & 9.85(15)  & 9.76(15)  & 9.07(15) 
			& 12.61(23) & 12.55(23) & 11.66(21) \\
	& 1 	& 49	& 9.93(13)  & 9.83(13)  & 9.06(13) 
			& 12.74(18) & 12.68(17) & 11.66(18) \\
	&$\ge2$ & 53 	& 9.750(92) & 9.650(90) & 9.040(97) 
			& 12.39(14) & 12.34(14) & 11.64(14) \\
2.875	& 0	& 29	& 8.64(16)  & 8.55(16)  & 7.89(19) 
			& 11.16(25) & 11.11(25) & 10.31(24) \\
	& 1	& 40	& 8.58(12)  & 8.50(12)  & 7.86(12) 
			& 11.02(17) & 10.97(17) & 10.15(18) \\
	&$\ge2$ & 59 	& 8.416(73) & 8.338(72) & 7.771(89) 
			& 10.68(10) & 10.633(99)& 10.02(12) \\
2.816	& 0	& 24	& 7.281(99) & 7.212(98) & 6.68(12)  
			& 9.32(15)  & 9.27(15)  & 8.63(16)  \\
	& 1	& 42	& 7.103(75) & 7.036(74) & 6.540(93) 
			& 9.06(12)  & 9.02(12)  & 8.41(12)  \\
	&$\ge2$ & 62	& 7.044(66) & 6.979(65) & 6.435(80) 
			& 8.964(91) & 8.924(91) & 8.22(11)  \\
2.751	& 0	& 28	& 5.878(70) & 5.822(69) & 5.381(66) 
			& 7.55(11)  & 7.52(11)  & 7.006(95) \\
	& 1	& 37	& 5.895(63) & 5.840(62) & 5.416(75) 
			& 7.542(96) & 7.507(95) & 7.10(11)  \\
	&$\ge2$ & 63	& 5.882(43) & 5.828(43) & 5.382(51) 
			& 7.491(61) & 7.456(62) & 6.920(65) \\
2.710	& 0	& 20	& 5.277(66) & 5.227(65) & 4.803(59) 
			& 6.750(90) & 6.720(90) & 6.185(97) \\
	& 1	& 30	& 5.229(48) & 5.179(47) & 4.825(73) 
			& 6.707(77) & 6.676(73) & 6.267(92) \\
	&$\ge2$ & 78	& 5.175(24) & 5.127(24) & 4.781(34) 
			& 6.615(34) & 6.585(34) & 6.161(45) \\
\hline\hline
\end{tabularx} \label{tab:scales} \end{table*}

To relate $\chi^{1/4}$ to physical scales, we use from Table~IX of 
Ref.~\cite{BC17} the relation $1/T_c=(2.2618\pm 0.0042)\,L_{10}$,
where $T_c$ is the SU(2) deconfinement temperature in lattice 
units. Propagating the statistical errors, we obtain from
Eqs.~(\ref{L10s1000}) and (\ref{L10s0100})
\begin{eqnarray} \label{sTc1000}
  \chi^{1/4}/T_c&=& 0.643\ (12)~~{\rm for}~~n_c=1000\,,\\ \label{sTc0100}
  \chi^{1/4}/T_c&=& 0.657\ (12)~~{\rm for}~~n_c=100\,.
\end{eqnarray}
In the literature $\chi^{1/4}$ for SU(2) LGT has been reported in units 
of the string tension $\sqrt{\sigma}$. The most accurate estimate of 
$T_c/\sqrt{\sigma}$ appears to be $T_c/\sqrt{\sigma}=0.7091\,(36)$ 
from Ref.~\cite{Lu04}, which is consistent with the earlier value 
$T_c/\sqrt{\sigma}=0.69\,(2)$ \cite{Fi93}. Using the former and error 
propagation our estimates (\ref{sTc1000}) and (\ref{sTc0100}) convert 
to
\begin{eqnarray} \label{soss1000}
  \chi^{1/4}/\sqrt{\sigma} &=& 0.4557\ (83)~~{\rm for}~~n_c=1000\,,
  \\ \label{soss0100}
  \chi^{1/4}/\sqrt{\sigma} &=& 0.4655\ (88)~~{\rm for}~~n_c=100\,.
\end{eqnarray}
In Table~\ref{tab_s1o4} we compile estimates of the literature. The last 
two columns report Gaussian difference tests obtained by comparing with 
our estimates (\ref{soss1000}) and (\ref{soss0100}). Both of our 
estimates are lower than each of the others, but this is not surprising 
since the value for the topological susceptibility goes down with 
increasing $n_c$. Our $n_c=100$ estimate of $\chi^{1/4}/\sqrt{\sigma}$ 
is statistically already consistent with all but one of the literature. 
That does not mean that it is a better estimate than that at $n_c=1000$, 
because the previous literature relied on rather small lattice sizes and 
$\beta$ values for which only small $n_c$ can be used. It may well be 
that $n_c=100$ is too small, and we suggest that our $n_c=1000$ results 
(\ref{sTc1000}) and (\ref{soss1000}) are the best. Although there is a 
danger of destroying real instantons when the value of $n_c$ is taken 
too large, there is no evidence for that happening in 
Figs~\ref{fig:c40b2p751} or~\ref{fig:c60b2p928}.

\section{Scales in topological sectors} \label{sec:scales}

For $\beta\ge 2.71$ we calculated cooling scales on the largest lattice 
in the topological sectors $Q^{1000}\le -2$, $Q^{1000}=-1$, $Q^{1000}=0$, 
$Q^{1000}=1$ and $Q^{1000}\ge 2$, and performed Student difference tests 
of each scale with itself on distinct topological sectors. 
No statistically significant discrepancies are encountered. 
In particular there are none when comparing the $Q^{1000}< 0$ with the 
$Q^{1000}>0$ scales. To increase the statistics for the $|Q^{1000}|\ne 
0$ sectors, we combined them into $|Q^{1000}|=1$ and $|Q^{1000}|\ge 2$. 
Together with the scales for $Q^{1000}=0$ their values are listed in 
Table~\ref{tab:scales}. The scales $L_7$ and $L_8$  as well as for 
$L_{10}$ and $L_{11}$ almost agree because the fluctuations of 
the operators $E_0$ and $E_1$ are strongly correlated and almost 
identical \cite{BC17}. So, they are combined in following.

\begin{figure}[t]
\caption{Histogram of $q$-values comparing cooling scales from
Table~\ref{tab:scales} across topological sectors.}
\includegraphics[width=\columnwidth]{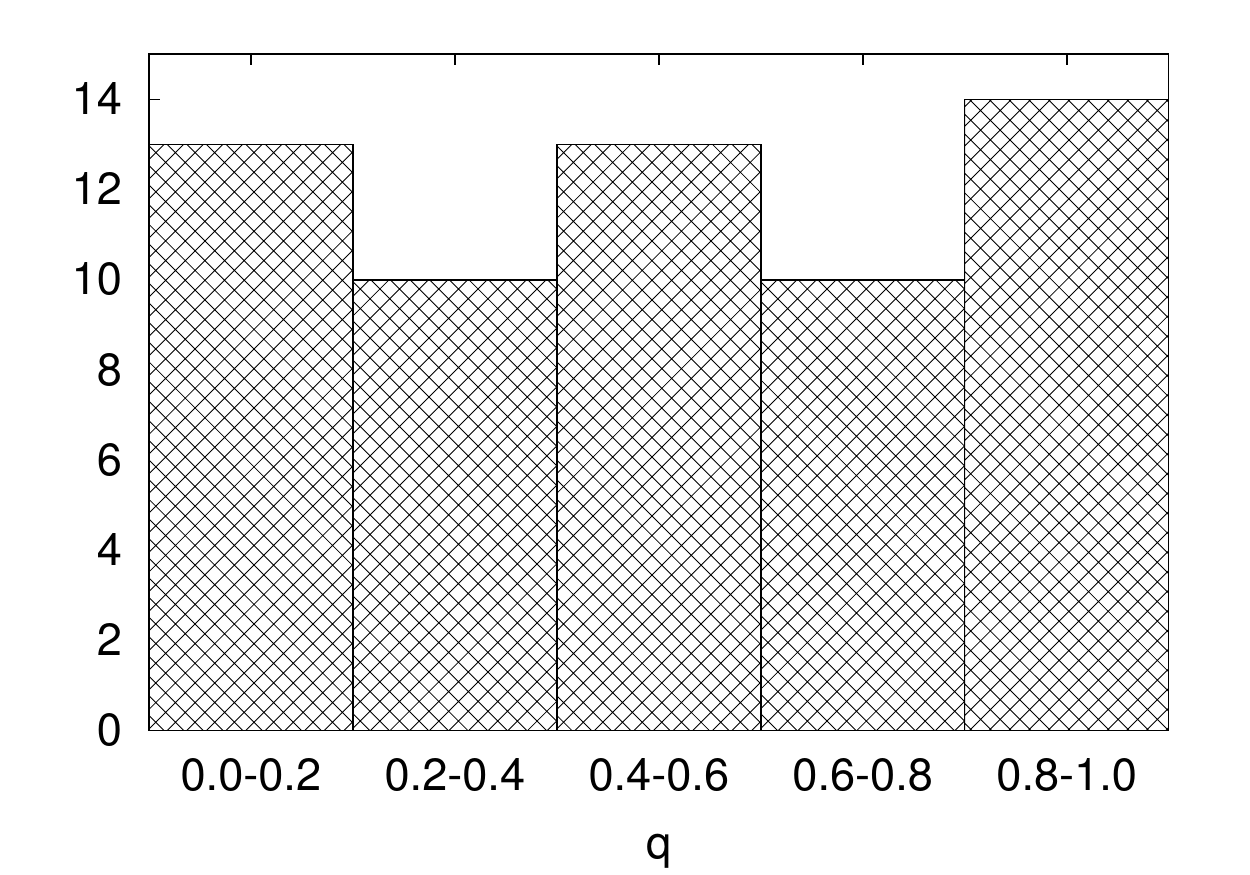}
\label{fig:qhistogram} \end{figure}

A histogram of the $q$ values of the remaining $4\times 15=60$ Student 
difference tests for the scales of Table~\ref{tab:scales} is shown
in Fig.~\ref{fig:qhistogram}. When the compared data are statistically 
independent, rely on the same estimator, and are drawn from a Gaussian 
distribution, the Student different tests return uniformly distributed 
random numbers $q$ in the range $0<q<1$, which is consistent with 
Fig.~\ref{fig:qhistogram}. Furthermore, their mean value comes out to 
be $\overline{q}=0.508\,(40)$ in agreement with the expected $0.5$. 
If there are still some residual correlations between our $q$-values, 
this would have decreased the error bar, because the number of 
independent $q$ would have been counted too high, while each of them
still fluctuates like a uniformly distributed random number in the 
interval (0,1). So, we find convincing evidence that the $1/V$ bias 
expected for our scales due to topological freezing disappears within 
our statistical noise.

\section{Summary and conclusions \label{sec:sum}}

Using standard cooling we calculated the topological charge of pure 
SU(2) LGT for larger lattices and $\beta$ values than it was done in 
the literature. For the first time they appear to be large enough to 
yield stable topological sectors. See Figs.~\ref{fig:c40b2p751}, 
\ref{fig:c60b2p928}, and~\ref{fig:s1o4}. From these data we obtain
the estimates (\ref{sTc1000}) to (\ref{soss0100}), which are 
surprisingly close to previous results of the literature listed 
in Table~\ref{tab_s1o4}. This may well be an accident, as the 
$n_c=1000$ versus $n_c=100$ fits of Fig.~\ref{fig:BAV} illustrate.

Within our statistical fluctuations we find no observable correlations 
between cooling scales (\ref{yi}) and topological charge sectors. Our 
number of statistically independent configurations is of a typical size 
as used for scale setting, e.g., \cite{L10,LS11}. So, our results
support that the problem of topological freezing only becomes serious 
when a much higher precision is targeted. 
\bigskip 

\centerline{\bf Acknowledgments} \smallskip

David Clarke was in part supported by the US Department of Energy 
(DOE) under contract DE-SC0010102. Our calculations used resources 
of the National Energy Research Scientific Computing Center (NERSC), 
a DOE Office of Science User Facility supported by the DOE under 
Contract DE-AC02-05CH11231.

\end{document}